\documentstyle[prd,aps,preprint]{revtex}

\tightenlines

\begin{document}

\title{Geodesically complete nondiagonal inhomogeneous cosmological
solutions \\
in dilatonic gravity with a stiff perfect fluid }
\draft
\author{Stoytcho S. Yazadjiev\footnote{E-mail:
yazad@phys.uni-sofia.bg}}
\address{Department of Theoretical Physics,  Faculty of Physics \\
Sofia University\\ 5 James Bourchier Boulevard \\1164 Sofia,
Bulgaria}
\maketitle
\begin{abstract}

New nondiagonal $G_{2}$ inhomogeneous  cosmological solutions are
presented in a wide range of scalar-tensor theories with a stiff
perfect fluid as a matter source. The solutions have no big-bang
singularity or any other curvature singularities. The dilaton
field and the fluid energy density are regular everywhere, too.
The geodesic completeness of the solutions is investigated.
\end{abstract}

\pacs{04.50.+h, 04.20.Jb, 98.80.Hw}



All versions of string theory and higher dimensional gravity
theories predict the existence of the dilaton field which
determines the gravitational "constant" as a variable quantity.
The existence of a scalar partner of the tensor graviton may have
a serious influence on the space-time structure and important
consequences for cosmology and astrophysics. A large amount of
research has been done in order to unveil the possible
cosmological significance of the dilaton \cite{GAS},
\cite{LS}-\cite{M}(and references therein). With a few exceptions
most of the cosmological studies within the scalar-tensor
theories were devoted to the homogeneous case. The homogeneous
models are good approximations of the present universe. There is,
however, no reason to assume that such a regular expansion is
also suitable for a description of the early universe. Moreover,
as is well known, the present universe is not exactly spacially
homogeneous. That is why it is necessary to study inhomogeneous
cosmological models. They allow us to investigate a number of
long standing questions regarding the occurrence  of
singularities, the behaviour of the solutions in the vicinity of a
singularity and the possibility of our universe arising from
generic initial data.

In this work we shall address the question of the occurrence of
singularities in inhomogeneous cosmologies within the framework of
scalar-tensor theories.

As is well known, most of the homogeneous models (both in general
relativity and in scalar-tensor theories) predict a universal
space-like big-bang singularity in a finite past. It was,
therefore, believed that this would be the usual singularity in
general. The inclusion of inhomogeneities  drastically changes
this point of view. There are inhomogeneous cosmological
solutions in general relativity which  have no bing-bang or any
other curvature singularity. The first such solution was
discovered by Senovilla in 1990 Ref.\cite{SEN}. Senovilla's
solution represents a cylindrically symmetric universe filled with
radiation. This solution has a diagonal metric and  is also
globally hyperbolic and geodesically complete \cite{CFJS}.
Senovilla's solution was generalized by Ruiz and Senovilla in
Ref.\cite{RS} where a large family of singularity-free diagonal
$G_{2}$ inhomogeneous perfect fluid solutions was found.
Nonsingular diagonal inhomegeneous solutions in general
relativity describing cylindrically symmetric universes filled
with stiff perfect fluid were found by Patel and Dadhich in
Ref.\cite{PATDAD}. Other examples of diagonal nonsingular
solutions in general relativity can be found in
Refs.\cite{PATDAD1}- \cite{LFJ1}.

In Ref.\cite{MM}, Mars found the first nondiagonal $G_{2}$
inhomogeneous cosmological solution of the Einstein equations
with stiff perfect fluid as a source. This solution is globally
hyperbolic and geodesically complete. Mars's solution was
generalized by Griffiths and Bicak in Ref.\cite{GB}.

Within the framework of scalar-tensor theories there are also
inhomogeneous cosmological solutions  without big-bang or any
other curvature singularity. In Ref.\cite{GIOV}, Giovannini
derived gravi-dilaton inhomogeneous cosmological solutions with
everywhere regular curvature invariants and bounded dilaton in
tree-level dilaton driven models. In a subsequent paper
\cite{GIOV1}, it was shown that these solutions describe
singularity-free dilaton driven cosmologies. A nondiagonal
inhomogeneous cosmological solution with regular curvature
invariants and unbounded dilaton in the tree level effective
string models was found by Pimentel \cite{PIM}. Very recently,
inhomogeneous cosmological solutions without any curvature
singularities were obtained by the author in a wide class of
scalar-tensor theories with stiff perfect fluid as a source
\cite{Y}.

In this work we take a further step upwards and  present new
nondiagonal $G_{2}$ inhomogeneous cosmological stiff perfect fluid
solutions with no curvature singularities in a wide range of
scalar-tensor theories.

Scalar-tensor theories (without a cosmological potential) are
described by the following action in Jordan (string) frame
\cite{W1},\cite{W2}:
\begin{eqnarray} \label{JFA}
S = {1\over 16\pi G_{*}} \int d^4x \sqrt{-g}\left(F(\Phi)R -
Z(\Phi) g^{\mu\nu}\partial_{\mu}\Phi
\partial_{\nu}\Phi  \right)  \\+
S_{m}\left[\Psi_{m};g_{\mu\nu}\right] .\nonumber
\end{eqnarray}

Here, $G_{*}$ is the bare gravitational constant and $R$ is the
Ricci scalar curvature with respect to the space-time metric
$g_{\mu\nu}$. The dynamics of the scalar field $\Phi$ depends on
the functions $F(\Phi)$ and $Z(\Phi)$. In order for the
gravitons  to carry positive energy the function $F(\Phi)$ must
be positive. The nonnegativity of the energy of the dilaton
requires that $2F(\Phi)Z(\Phi)+ 3[dF(\Phi)/d\Phi ]^2\ge 0$. The
action of matter depends on the material fields $\Psi_{m}$ and the
space-time metric $g_{\mu\nu}$ but does not involve the scalar
field $\Phi$ in order for the weak equivalence principle to be
satisfied.

As a matter source we consider a stiff perfect fluid with
equation of state $p=\rho$.

The general form of the solutions is given by

\begin{eqnarray}
d{s}^2= F^{-1}(\Phi(t))\left[ e^{\gamma a^2r^2}\cosh(2at)(-dt^2 +
dr^2) \right. \nonumber \\ \left. + \,\, r^2\cosh(2at)d\phi^2 +
{1\over \cosh(2at)}(dz + ar^2d\phi)^2 \right] \nonumber ,
\end{eqnarray}

\begin{eqnarray}\label{STS}
8\pi G_{*}\rho = f(\lambda) {a^2 (\gamma - 1)F^3(\Phi(t))
e^{-\gamma a^2 r^2} \over \cosh(2at)}  ,
\end{eqnarray}

\begin{eqnarray}
u_{\mu} =F^{-1/2}(\Phi(t)) e^{(1/2)\gamma a^2 r^2}
\cosh^{1/2}(2at)\,\delta^{0}_{\mu}  \nonumber .
\end{eqnarray}

The solution depends on three parameters - $a$, $\gamma$ ($\gamma
>1 $) and $\lambda$. The range of the coordinates is

\begin{equation}
-\infty <t, z < \infty ,\,\,\, 0\leq r < \infty , \,\,\, 0\leq
\phi \leq 2\pi .
\end{equation}

The explicit form of the functions $\Phi(t)$ and $f(\lambda)$,
and the range of the parameter $\lambda$  depend on the
particular scalar tensor theory. These solutions can be
generated\footnote{Some of the solutions were first obtained by
solving the corresponding system of partial differential
equations for nondiagonal $G_{2}$ cosmologies \cite{Y1}.} from the
general relativistic Mars's solution  \cite{MM} using the solution
generating methods developed in Ref.\cite{Y}.

Below we consider the explicit form of the general solution for
some particular scalar-tensor theories.

\subsection{Barker's theory}

Barker's theory is described by the functions $F(\Phi)=\Phi$ and
$Z(\Phi)= (4-3\Phi)/2\Phi(\Phi - 1)$.

In the case of Barker's theory the explicit forms of the functions
$\Phi(t)$ and $f(\lambda)$ are:

\begin{eqnarray}
\Phi^{-1}(t)& =& 1 -\lambda  \cos^2\left(a\sqrt{\gamma
-1}t\right),
\\
f(\lambda) &=& 1-\lambda
\end{eqnarray}

where the range of $\lambda$ is  $0< \lambda < 1$. This range can
be extended to $0\le \lambda \le 1$. For $\lambda=0$ and
$\lambda=1$ we obtain the Mars's solution and gravi-dilaton vacuum
solution, respectively. That is why we consider  only $0<\lambda
<1$. It should be noted that the range of the parameter $\lambda$
is crucial for the curvature invariants. It is easy to see that
the gravi-dilaton vacuum solution corresponding to $\lambda=1$
has divergent curvature invarinats because of the conformal
factor $\Phi^{-1}(t)= \sin^2\left(a\sqrt{\gamma -1}t\right)$.

\subsection{Brans-Dicke theory}

Brans-Dicke theory is described by the functions $F(\Phi)=\Phi$
and $Z(\Phi) = \omega /\Phi$ where $\omega$ is a constant
parameter. Here we consider the case $\omega >-3/2$. The explicit
form of the functions $\Phi(t)$ and $f(\lambda)$ in the
Brans-Dicke case is the following:

\begin{eqnarray}
\Phi^{-1/2}(t) &=& \lambda \exp\left(a\sqrt{\gamma -1\over 3+
2\omega}\,t\right) \nonumber \\ &+& (1-\lambda)\exp\left( -
\,a\sqrt{\gamma
-1\over 3+2\omega}\,t\right), \\
f(\lambda) &=& 4\lambda(1-\lambda).
\end{eqnarray}

Here the range of the parameter $\lambda$ is $0< \lambda <1$. The
solution exists for $\lambda=0$ and $\lambda=1$, too. In these
cases, however, we obtain a gravi-dilaton vacuum solution which is
just the Pimentel's solution \cite{PIM}. That is the reason we do
not consider these limiting values of $\lambda$. The solution is
invariant under the trasformations $\lambda \longleftrightarrow
1-\lambda$ and $t\longleftrightarrow -t$. In this generalized
sense, we can say that the solution is even in time.

\subsection{Theory with "conformal" coupling }

The theory with "conformal" coupling is described by the functions
$F(\Phi) = 1 - {1\over 6}\Phi^2 $ and $Z(\Phi)= 1$. In this case
we have:

\begin{eqnarray}
F^{-1}(\Phi(t)) &=& 1 +
4\lambda(1-\lambda)\sinh^2\left(a\sqrt{\gamma
-1\over 3}\, t\right), \\
f(\lambda) &=& (1-2\lambda)^2 .
\end{eqnarray}

The range of the parameter $\lambda$ is $0<\lambda \le 1/2$. For
$\lambda=1/2$ we obtain a gravi-dilaton vacuum solution which is
well-behaved and can be included as a limiting case.

\subsection{$Z(\Phi)= {(\Omega^2 - 3\Phi)/ 2\Phi^2}$ theory}

Here we consider the scalar-tensor theory described by the
functions $F(\Phi)= \Phi$ and $Z(\Phi)= (\Omega^2 - 3\Phi)/
2\Phi^2$ where $\Omega>0$. The explicit forms of $\Phi(t)$  and
$f(\lambda)$ are:

\begin{eqnarray}
\Phi^{-1}(t) &=& \left(1 + {1 \over \Omega}a\sqrt{\gamma -1}t
\right)^2 + \lambda , \\
f(\lambda) &=& \lambda .
\end{eqnarray}

Here, the range of the parameter is $0< \lambda < \infty$.

\subsection{$Z(\Phi)= {1\over 2}(\Phi^2 - 3\Phi + 3)/\Phi(\Phi -1)$ theory
}

The theory with $F(\Phi)=\Phi$ and $Z(\Phi)= {1\over 2}(\Phi^2 -
3\Phi + 3)/\Phi(\Phi -1)$ possesses the following solution:

\begin{eqnarray}
\Phi^{-1}(t) &=& {\lambda^2 \over \lambda^2 +
(1-\lambda^2)\sin^2(\lambda a\sqrt{\gamma -1} t) } ,\\
f(\lambda) &=& \lambda^2.
\end{eqnarray}

In order for the dilaton field  in this  solution to  have
positive energy  we should restrict the range of the parameter
$\lambda$ to $0<\lambda<1$.

Using the solution generating methods developed in Ref.\cite{Y} we
can generate nondiagonal $G_{2}$ inhomogeneous cosmological
solutions in many other scalar-tensor theories different from
those considered above. However, the solutions we have presented
here are expressed in a closed analytic form and they are also
representative and cover a wide range of the possible behaviors of
the scalar-tensor solutions which can be generated from Mars's
solution.

Let us consider the main properties of the found solutions. The
metric functions, the gravitational scalar (the dilaton) and the
fluid energy density are everywhere regular. The space-times
described by our solutions have no big-bang nor any other
curvature singularity - the curvature invariants $I_{1}=
C_{\mu\nu\alpha\beta}C^{\mu\nu\alpha\beta}$, $I_{2}=
R_{\mu\nu}R^{\mu\nu}$, and $I_{3} = R^2$ are regular everywhere.
The solution possesses a two dimensional abelian group of
isometries inherited from the seed Mars's solution and generated
by the Killing vectors $\partial/\partial z$ and
$\partial/\partial \phi$. In addition, the metrics have a well
defined axis of symmetry and the elementary flatness condition
\cite{KSHMC} is satisfied. Since the presented solutions are
conformally related to the Mars's solution, the spacetimes
descibed by them are globally hyperbolic. In fact, the global
hyperbolicity can be proved independently as a consequence of the
poof of the geodesic completeness presented below.

The existence of two Killing vectors gives rise to two constants
of motion along the geodesics:

\begin{eqnarray}\label{FIN}
K= F^{-1}(\Phi(t)) \nonumber  \\\times
\left[\cosh(2at)r^2{d\phi\over ds} + {ar^2\over
\cosh(2at)}\big({dz\over ds} + ar^2 {d\phi \over ds}\big) \right], \\
L= {F^{-1}(\Phi(t))\over \cosh(2at)} ({dz\over ds} + ar^2 {d\phi
\over ds}\nonumber ).
\end{eqnarray}

The affinely parameterized causal geodesics satisfy

\begin{eqnarray}\label{AFP}
F^{-1}(\Phi(t))\{e^{\gamma a^2r^2}\cosh(2at) [({dt\over ds })^2 -
({dr\over ds })^2 ]  \nonumber  \\ - {L^2 \cosh(2at)\over
F^{-2}(\Phi(t))} - {(K -Lar^2)^2\over r^2 F^{-2}(\Phi(t))
\cosh(2at)}\} = \epsilon
\end{eqnarray}

where $\epsilon =0$ and $1$ for null and timelike geodesics,
respectively. Taking into account (\ref{FIN}) and  (\ref{AFP}) the
geodesic equations for $t$ and $r$ can be written in the
following form:

\begin{eqnarray}
{d\over ds }\left( F^{-1}(\Phi(t))e^{\gamma a^2
r^2}\cosh(2at){dt\over ds } \right) \\ = F(\Phi(t))e^{-\gamma a^2
r^2} \cosh^{-1}(2at)M(t,r)\partial_{t}M(t,r) ,
\nonumber \\
{d\over ds }\left( F^{-1}(\Phi(t))e^{\gamma a^2
r^2}\cosh(2at){dr\over ds } \right)  \\ = - F(\Phi(t))e^{-\gamma
a^2 r^2} \cosh^{-1}(2at)M(t,r)\partial_{r}M(t,r) ,\nonumber
\end{eqnarray}

where

\begin{eqnarray}
M(t,r) =  F^{-1/2}(\Phi(t))e^{(1/2)\gamma a^2 r^2}\cosh^{1/2}(2at)
\nonumber \\ \times \left[\epsilon  + {L^2 \cosh(2at)\over
F^{-1}(\Phi(t))} + {(K - Lar^2)^2 \over r^2 F^{-1}(\Phi(t))
\cosh(2at)} \right]^{1/2}.
\end{eqnarray}

To demonstrate the geodesic completeness of our metric, we have to
show that all non-spacelike (i.e. causal) geodesics can be
extended to arbitrary values of the affine parameter. We shall
consider only future directed geodesics. The past directed
geodesics can be treated analogously.

First we consider null geodesics with $K=L=0$. For them we have
${dt\over ds}=|{dr\over ds}|$ and

\begin{equation}
{d\over ds }\left( F^{-1}(\Phi(t))e^{\gamma a^2
r^2}\cosh(2at){dt\over ds } \right) \\ =0.
\end{equation}

After integrating we obtain

\begin{equation}
{dt\over ds} = C \left(F^{-1}(\Phi(t))e^{\gamma a^2
r^2}\cosh(2at)\right)^{-1}
\end{equation}

where $C>0$ is a constant. Taking into account that for each of
our solutions there exists a constant $B$ such that

\begin{equation}
0 <B\le F^{-1}(\Phi(t))
\end{equation}

for arbitrary values of $t$ and fixed parameter $\lambda$, we
obtain

\begin{equation}
{dt\over ds}=|{dr\over ds}|\le {C\over B}.
\end{equation}

Therefore the geodesics under consideration are complete.

Now let us turn to the general case when at least one of the
constants $\epsilon$, $K$ or $L$ is different from zero. Here we
shall use a method similar to that  for diagonal metrics
described in Ref.\cite{CFJS}. Let us parameterize $dt\over ds$ and
$dr\over ds$  by writing :

\begin{eqnarray}\label{PGE1}
{dt\over ds} = {F(\Phi(t))e^{-\gamma a^2r^2}\over \cosh(2at)}
M(t,r)\cosh(\upsilon), \\
{dr\over ds} = {F(\Phi(t))e^{-\gamma a^2r^2}\over \cosh(2at)}
M(t,r)\sinh(\upsilon).
\end{eqnarray}

Substituting these expressions in the equations for $t$ and $r$
we obtain

\begin{eqnarray}\label{PGE2}
{d\upsilon \over ds } = -{F(\Phi(t))e^{-\gamma a^2r^2}\over
\cosh(2at)}\left[\partial_{t}M(t,r)\sinh(\upsilon) \right.
\nonumber \\ \left. +
\partial_{r}M(t,r)\cosh(\upsilon) \right]
\end{eqnarray}

or equivalently

\begin{eqnarray}\label{EV}
{d\upsilon\over ds } = - {1\over 2 M(t,r)} \{
\Gamma_{+}(t,r)e^{\upsilon}  + \Gamma_{-}(t,r)e^{-\upsilon}\}
\end{eqnarray}

where

\begin{eqnarray}\label{GPM}
\Gamma_{+}(t,r) = \epsilon \left[ a\tanh(2at) + {1\over
2}{d\ln[F^{-1}(\Phi(t))] \over dt} + \gamma a^2 r \right] \nonumber \\
+ { (K - Lar^2)^2 \over r^2 F^{-1}(\Phi(t))\cosh(2at)}
\left[\gamma a^2 r - {1\over r} - {2Lar \over K - Lar^2} \right]
\nonumber \\ + {L^2 \cosh(2at)\over F^{-1}(\Phi(t))}
\left[2a\tanh(2at) + \gamma a^2 r  \right], \\
\Gamma_{-}(t,r) = \epsilon \left[- a\tanh(2at) - {1\over
2}{d\ln[F^{-1}(\Phi(t))] \over dt} + \gamma a^2 r \right] \nonumber \\
+ { (K - Lar^2)^2 \over r^2 F^{-1}(\Phi(t))\cosh(2at)}
\left[\gamma a^2 r - {1\over r} - {2Lar \over K - Lar^2} \right]
\nonumber \\ + {L^2 \cosh(2at)\over F^{-1}(\Phi(t))}
\left[-2a\tanh(2at) + \gamma a^2 r  \right].\nonumber
\end{eqnarray}

In order for the geodesics to be complete ${dt \over ds }$ and
${dr\over ds}$ have to remain finite for finite values of the
affine parameter. In fact, it is sufficient  to consider only
${dt\over ds}$, since ${dr\over ds}$ and ${dt\over ds}$ are
related via (\ref{AFP}). The derivatives ${d\phi \over ds}$ and
${dz \over ds}$ are regular functions of $t$ and $r$, and the
only problem we could have appear when $r$ approaches $r=0$ for
$K\ne 0$. We shall show, however, that $r$ cannot become zero for
$K\ne 0$.

First we consider geodesics with increasing $r$ (i.e. $\upsilon
> 0$). In this case it is not difficult to see that the term

\begin{equation}
{F(\Phi(t))e^{-\gamma a^2r^2}\over \cosh(2at)} M(t,r)
\end{equation}

in eqn. (\ref{PGE1}) can not become singular (for increasing $r$).
Therefore, $dt \over ds $ could become singular only for
$\upsilon$. We shall show, however, that $\upsilon$ can not
become singular for finite values of the affine parameter. For
increasing $r$,  $\upsilon$ cannot diverge  since for large $t$
(large $r$ ) the derivative ${d\upsilon\over ds}$ becomes
negative. Indeed, for all exact solutions presented here, there
exists a constant $B_{1}>0$ such that

\begin{equation}
\mid{d\ln[F^{-1}(\Phi(t))]\over dt }\mid< B_{1}
\end{equation}

for arbitrary $t$ and fixed $\lambda$.

Therefore, as can be seen from eqns. (\ref{GPM}), the terms
associated with the constant $\varepsilon$, $K$ and $L$ are all
positive for large values of $t$.As a consequence we obtain that
the functions $\Gamma_{+}(t,r)$ and $\Gamma_{-}(t,r)$ are
positive, i.e. ${d\upsilon\over ds}<0$ for large $t$(large
$r$).\footnote{In fact, the function $\Gamma_{-}(t,r)$ is
exponentially small compared with $\Gamma_{+}(t,r)$ and may not
be considered.}

In the second case,when $r$ decreases ($\upsilon<0$), the problem
comes from  $r=0$ when $K \ne 0$. The geodesics with $K=0$ can
reach the axis $r=0$ without problems and then continue with
${dr\over ds}>0$ ($\upsilon>0$). When $K\ne 0$, $\upsilon$ cannot
diverge for finite values of the affine parameter. This follows
from the fact that the derivative ${d\upsilon\over ds}$ becomes
positive for small $r$ (large $t$) as can be seen from Eqns.
(\ref{GPM}) and (\ref{EV}), taking into account that
$\Gamma_{+}(t,r)$ is exponentially suppressed compared with
$\Gamma_{-}(t,r)$. The positiveness of the derivative
${d\upsilon\over ds}$ when the geodesics are close to the axis
$r=0$ prevents the radial coordinate from collapsing  too quickly
and reaching the axis. The fact that  $r$ can not become zero for
$K\ne 0$ may be seen more explicitly as follows. When $r$
approaches zero the dominant term is that associated with $K$ and
the other terms can be ignored. So, for small $r$ the geodesics
behaves as null geodesics with $L=0$:

\begin{eqnarray}
{dt\over ds} = {F(\Phi(t))e^{-\gamma a^2r^2}\over \cosh(2at)}
M(r)\cosh(\upsilon), \\
{dr\over ds} = {F(\Phi(t))e^{-\gamma a^2r^2}\over \cosh(2at)}
M(r)\sinh(\upsilon), \\
{d\upsilon \over ds } = -{F(\Phi(t))e^{-\gamma a^2r^2}\over
\cosh(2at)}
\partial_{r}M(r)\cosh(\upsilon)
\end{eqnarray}

where $M(r)= {|K|\over r}e^{(1/2)\gamma a^2r^2}$. Hence, we obtain
the orbit equation

\begin{equation}
{dr\over d\upsilon} = -{M(r)\over \partial_{r}M(r)}
\tanh(\upsilon).
\end{equation}

Integrating, we have

\begin{equation}
e^{-(1/2)\gamma a^2r^2}r = C_{1}\cosh(\upsilon)
\end{equation}

where $C_{1}>0$ is  a constant. Since $\cosh(\upsilon)\ge 1$, $r$
can not become zero.

From the proof of the geodesic completeness it follows that every
maximally extended null geodesic intersects any of the
hypersurfaces $t=const$. According to \cite{Geroch}, this a
sufficient condition that the hypersurfaces $t=const$ are global
Cauchy surfaces. Therefore, the solutions are globally hyperbolic.

We have explicitly proven the geodesic completeness of the
solutions using their particular properties. The geodesic
completeness can be proved independently by considering the
solutions from a more general point of view. In
Ref.\cite{LFJ2}(see also Ref.\cite{LFJ3}), Fernandez-Jambrina
presented a general theorem providing  wide sufficient conditions
for an orthogonally transitive cylindrical space-time to be
geodesically complete. It can be verified that the solutions
presented here satisfy all conditions in the Fernandez-Jambrina's
theorem and therefore they are geodesically complete.

New diagonal solutions can be obtained  from (\ref{STS}) as a
limiting case. Taking $a\rightarrow 0$ and keeping
$a^2\gamma=\beta$ fixed, we obtain the following diagonal
inhomogeneous cosmological scalar-tensor solutions:

\begin{eqnarray}
ds^2 &=& F^{-1}(\Phi(t)) \left[e^{\beta r^2}(-dt^2 + dr^2) +
r^2d\phi^2 + dz^2 \right], \nonumber\\
8\pi G_{*}\rho &=& \beta f(\lambda)e^{-\beta r^2}F^{3}(\Phi(t)) ,\\
u_{\mu}&=& F^{-1/2}(\Phi(t)) e^{(1/2)\beta
r^2}\delta^{0}_{\mu}\nonumber
\end{eqnarray}

where $a\sqrt{\gamma -1}$ should be replaced by $\sqrt{\beta}$ in
the explicit formulas for $F^{-1}(\Phi(t))$.

We have proven that the solutions presented in the present paper
are geodesically complete. This result is not in contradiction
with the well-known singularity theorems because in our case the
strong energy condition is violated in the Jordan frame. This can
be explicitly seen by calculating the components of the Ricci
tensor. All components are bounded except for $R_{tr}= -r \gamma
a^2 \partial_{t} \ln\{F[\Phi(t)]\}$. Therefore, for large enough
$r$, one can always find timelike and null vectors
$\upsilon^{\mu}$ such that
$R_{\mu\nu}\upsilon^{\mu}\upsilon^{\nu}<0$ i.e. the strong energy
condition is violated. However, the situation is different in the
Einstein frame. The Einstein frame metric $g^{E}_{\mu\nu}$ is just
the Mars's metric and it is geodesically complete as we have
already mentioned. Since the energy conditions are satisfied in
the Einstein frame it remains to see which other conditions of the
singularity theorems are violated. The space-time described by the
metric $g^{E}_{\mu\nu}$ does not contain closed trapped surfaces.
In order to prove this we will employ the techniques of
differential geometry described in Refs.\cite{SEN2} and
\cite{SEN3}. Let us consider a closed spacelike surface $\cal{S}$
and suppose that it is trapped. Since the surface is compact it
must have a point $q$ where $r$ reaches its maximum. Let us denote
$r_{max}=R$ on a constant time hypersurface $t=T$. For the traces
of both null second fundamental forms at $q$, it can be shown
that (see Refs.\cite{SEN2} and\cite{SEN3})

\begin{eqnarray}
K^{+}_{\cal{S}}|_{q}\ge {e^{-(1/2)\gamma a^2R^2}\over
\sqrt{2}R \cosh^{1/2}(2aT)}>0 \nonumber , \\
K^{-}_{\cal{S}}|_{q}\le - {e^{-(1/2)\gamma a^2R^2}\over \sqrt{2}R
\cosh^{1/2}(2aT)}<0 .
\end{eqnarray}

The traces have opposite signs so that there are no trapped
surfaces.

Our solutions are stiff perfect fluid cosmologies and, therefore,
the natural question which arises is what happens if the fluid is
not stiff. In this case,however, the situation is much more
complicated. In contrary to the stiff fluid case, the
dilaton-matter sector does not posses nontrivial symmetries which
allow us to generate new solutions from known ones.The only way
to find exact solutions is to attack directly the corresponding
system of coupled partial differential equations. This question
is currently under investigation.

Summarizing, in this work we have presented new nondiagonal
$G_{2}$ inhomogeneous stiff perfect fluid cosmological solutions
in a wide range of scalar-tensor theories. The found solutions
have no big-bang nor any other curvature singularity. The
gravitational scalar (dilaton) and fluid energy density
(pressure) are regular everywhere, too. Moreover, the solutions
are globally hyperbolic and geodesically complete. To the best of
our knowledge, these solutions are the first examples  of
nonsingular $G_{2}$ inhomogeneous perfect fluid scalar-tensor
cosmologies with a nondiagonal metric.

\par \indent
I would like to thank V. Rizov for discussions and especially L.
Fernandez-Jambrina for his valuable comments on the geodesic
completeness of the orthogonally transitive cylindrical
spacetimes. My thanks also go to J. Senovilla for sending me some
valuable papers. This work was supported in part by Sofia
University Grant No 459/2001.




\begin{references}

\bibitem{GAS} M. Gasperini's web page,
http://www.to.infn.it{\,}${\tilde{}}${\,}gasperin

\bibitem{LS}{D. La}, {P. Steinhardt}, Phys. Rev. Lett. {\bf 62},
376 (1989)

\bibitem{PSS} {L. Pimentel}, {J. Stein-Schabes}, Phys. Lett. {\bf B
216}, 25 (1989)


\bibitem{BM} {J. Barrow}, {K. Maeda}, Nucl. Phys. {\bf B 341}, 294
(1990)


\bibitem{B} {J. Barrow},  Phys. Rev. {\bf D 47}, 5329 (1993)



\bibitem{BMIM} {J. Barrow}, {J. Mimoso}, Phys. Rev. {\bf D 50}, 3746
(1994)

\bibitem{KW3} {C. Will}, {P. Steinhardt},  Phys.Rev. {\bf D 52}, 628
(1995)

\bibitem{MW1} {J. Mimoso}, {D. Wands}, Phys. Rev. {\bf D 51}, 477
(1995)

\bibitem{MW2} {J. Mimoso}, {D. Wands}, Phys. Rev. {\bf D 52}, 5612
(1995)


\bibitem{BPAR} {J. Barrow}, {P. Parsons }, Phys. Rev. {\bf D 55},
1906 (1997)

\bibitem{BMAR} {O. Bertolami}, {P. Martins}, Phys. Rev. {\bf D 61},
064007-1 (2000)

\bibitem{BPIET} {N. Bartolo}, {M. Pietroni}, Phys. Rev. {\bf D 61}, 023518
(2000)


\bibitem{FP}{G. Esposito-Farese}, {D. Polarski}, Phys. Rev. {\bf D
63}, 063504 (2001)

\bibitem {M}{J. Morris}, Class.Quant.Grav. {\bf 18}  2977 (2001)



\bibitem{SEN} J. Senovilla, Phys. Rev. Lett. {\bf 64}, 2219 (1990)

\bibitem{CFJS} F. Chinea, L. Fernandez-Jambrina, J. Senovilla,
Phys. Rev {\bf D45}, 481 (1992)


\bibitem{RS}E.Ruiz, J. Senovilla, Phys.Rev.{\bf D45}, 1995 (1992)



\bibitem{PATDAD} L. Patel, N. Dadhich, Report No. IUCAA-1/93;
gr-qc/ 9302001


\bibitem{PATDAD1} L. Patel, N. Dadhich, Class. Quant. Grav. {\bf 10}, L85
(1993)

\bibitem{DPT1} N. Dadhich, L. Patel, R Tikekar, Current Science {\bf 65},
694 (1993)

\bibitem{SEN1} J. Senovilla, Phys. Rev. {\bf D53}, 1799 (1996)


\bibitem{ND1} N. Dadhich,  J. Astrophys. Astr. {\bf 18}, 343 (1997)

\bibitem{NR1} N. Dadhich,  A. Raychaudhuri, Mod. Phys. Lett {\bf A14},
2135 (1999).

\bibitem{MMDS} M. Mars, Class. Quant. Grav.{\bf 12}, 2831 (1995)

\bibitem{LFJ1} L. Fernandez-Jambrina, Class. Quant. Grav.{\bf 14},
3407 (1997)


\bibitem{MM} M. Mars, Phys. Rev. {\bf D51}, R3989 (1995)

\bibitem{GB} J. Griffiths, J. Bicak, Class. Quant. Grav.{\bf 12},
L81 (1995)


\bibitem{GIOV} M. Giovannini, Phys. Rev. {\bf D57}, 7223 (1998)

\bibitem{GIOV1} M. Giovannini, Phys. Rev. {\bf D59}, 083511 (1999)

\bibitem{PIM} L. Pimentel, Mod. Phys. Lett {\bf A}, {\bf Vol 14}, 43
(1999)

\bibitem{Y} S. Yazadjiev,  Phys. Rev. {\bf D65}, 084023 (2002)


\bibitem {W1} C. Will, The Confrontation between
 General Relativity and Experiment,Living Rev.Rel. 4 (2001) 4;
 E-print gr-qc/0103036


\bibitem{W2} C. Will {\em Theory and experiment in graviattional
physics} (Cambridge University Press, Cambridge, 1993)


\bibitem{Y1} S. Yazadjiev, Talk given at the Meeting of the
researchers in physics, Sofia University, Bulgaria,  2002

\bibitem{KSHMC}D. Kramer, H. Stephani, E. Herlt, M. MacCallum, {\em Exact
Solutions of Einstein's Field Equations}(Cambridge University
Press, Cambridge, 1980)

\bibitem{Geroch} R. Geroch, J. Math. Phys. {\bf 11}, 437 (1970)


\bibitem{LFJ2} L. Fernandez-Jambrina, J. Math. Phys. {\bf 40},
4028 (1999)


\bibitem{LFJ3}  L. Fernandez-Jambrina, L. M. Gonzalez-Romero,
Class.Quant.Grav. {\bf 16}, 953, (1999)

\bibitem{SEN2} J. Senovilla, Gen. Relativ. Gravit. {\bf 30}, 701
(1998)

\bibitem{SEN3} J. Senovilla, "Trapped surfaces, horizons, and exact
solutions in higher dimensions", hep-th/0204005

\end{references}
\end{document}